\documentstyle[11pt,aaspp4,psfig]{article}
\lefthead{Gallart et al. \etal}
\righthead{Radial velocity of Phoenix}

\newcommand\kms{${\rm km~s}^{-1}$~}

\begin{document}

\title{Radial velocity of the Phoenix dwarf galaxy: linking stars and HI 
gas\altaffilmark{1}}

\author{C. Gallart\altaffilmark{2,3}, D. Mart\'\i nez-Delgado\altaffilmark{4},
M.A. G\'omez-Flechoso\altaffilmark{5} \& M. Mateo\altaffilmark{6} }

Subject Headings:  galaxies: individual (Phoenix); galaxies: stellar 
content; galaxies: HI content.

\altaffiltext{1}{Based on observations collected in visitor mode with 
the VLT UT1, ANTU, at the European Southern Observatory, Chile}
\altaffiltext{2}{Universidad de Chile. Departamento de Astronom\'\i a. 
Casilla 36-D. Las Condes. Santiago. Chile.}
\altaffiltext{3}{Andes Prize Fellow, Yale University. Department of 
Astronomy. P.O. Box. 208101. New Haven, CT 06520-8101. USA.} 
\altaffiltext{4}{Instituto de Astrof\'\i sica de Canarias, E-38200 
La Laguna, Canary Islands, Spain} 
\altaffiltext{5}{Observatoire de Gen\`eve, CH-1290 Sauverny, Switzerland}
\altaffiltext{6}{Department of Astronomy, 821 Dennison Building, 
University of Michigan, Ann Arbor, Michigan 48109, USA}

\newpage

\begin{abstract}

We present the first radial velocity measurement of the stellar 
component of the Local Group dwarf galaxy Phoenix, using FORS1 at the 
VLT UT1 (ANTU) telescope. From the spectra of 31 RGB stars, we derive 
an heliocentric optical radial 
velocity of Phoenix $V_{\odot}=-52\pm 6$ \kms. On the basis 
of this velocity, and taking into account the results of a series of
semi-analytical and numerical simulations, we discuss the possible 
association of the HI clouds observed in the Phoenix vicinity. We 
conclude that the characteristics of the HI cloud with heliocentric 
velocity --23 \kms are consistent with this gas having been associated with
Phoenix in the past, and lost by the galaxy after the last event of
star formation in the galaxy, about 100 Myr ago. Two possible scenarios 
are discussed: the ejection of the gas by the energy released by 
the SNe produced in that last event of star formation, and a ram-pressure
stripping scenario. We derive that the kinetic energy necessary to eject 
the gas is $E_{SNe} \sim 2 \times 10^{51}$ erg, and that the number of 
SNe necessary to transfer this amount of kinetic energy to the gas cloud is 
$\sim 20$. This is consistent with the number of SNe expected for the 
last event of star formation in Phoenix, according to 
the star formation history derived by Mart\'\i nez-Delgado et al. (1999).  
The drawback of this scenario is the regular appearance of 
the HI cloud and its anisotropic distribution with respect to the stellar 
component. Another possibility is that the HI gas
was stripped as a consequence of ram--pressure by the intergalactic 
medium. In our simulations, the 
structure of the gas remains quite smooth as it is stripped from Phoenix, 
keeping a distribution similar to that of the observed HI cloud. 
Both in the SNe ejection case and in the ram-pressure sweeping scenario, 
the distances and relative velocities imply that the HI cloud is not 
gravitationally bound to Phoenix, since this would require a Phoenix total 
mass about an order of magnitude larger than its total estimated mass.
Finally, we discuss the possibility that Phoenix may be a bound Milky Way 
satellite. The minimum required mass of the Milky Way for Phoenix to be bound  
is $M_{MW}(<450 {\rm kpc}) \ge 1.2 \times 10^{12}$ M$_{\odot}$ which 
comfortably fits within most current estimates.

\end{abstract}

\keywords{galaxies:individual (Phoenix), galaxies: dwarf, 
galaxies:evolution, ISM: HI, Local Group}

\section{Introduction} \label{intro}

At the low end of the mass and luminosity
distributions, dwarf galaxies are the most common type of galaxies in
the Universe. They are usually divided in two groups: dwarf irregular
(dIrr) and dwarf spheroidal (dSph) galaxies. In the simplest scheme, the key
differences between the two groups is that dIrr have large amounts of
gas and substantial current star formation activity, while dSph lack
gas and recent star formation. However, this picture has been shown to
be too simplistic, and the distinction between the two classes is
not always straightforward.  There are dIrr galaxies, like LGS~3 
(Aparicio, Gallart \& Bertelli 1997b; Young \& Lo 1998) and
Pegasus (Aparicio, Gallart \& Bertelli 1997a; Lo,
Sargent \& Young 1993), with substantial amounts of gas but
apparently low star formation rate. Other galaxies, like 
Fornax and Leo~I, which are traditionally considered dSph, have 
relatively recent (few hundred Myr old) star formation (Stetson, 
Hesser \& Smecker-Hane 1998; Gallart et al. 1999a,b). Finally, there are 
some interesting cases where gas, not centered in the optical component 
of the galaxy, is or may be associated with dSph 
systems: Sculptor (Carignan et al. 1998), Tucana (Oosterlo, Da Costa \& 
Staveley-Smith 1996),  And III, And V, Leo I and Sextans (Blitz \& 
Robishaw 2000). In the case 
of Sculptor, the kinematic association of the galaxy with two HI clouds 
located symmetrically about the galaxy center has been confirmed by 
Carignan et al. (1998). 

From the theoretical point of view, the relationship between dIrr and dSph 
galaxies, and their possible evolutionary links have been profusely 
discussed
in the literature (e.g. Einasto et al. 1974; Lin \& Faber 1983; Dekel \& 
Silk 1986; Silk, Wyse \& Shields 1987; Davies \& Phillips 1988; Mac Low 
\& Ferrara 1999; Ferrara \& Tolstoy 2000). The fact that the same basic 
relations and overall structure are shared by dIrr and dSph (Bingelli 1993)
seem to point to a common origin. On the other hand, the morphological 
segregation of dwarfs, with dSph in general close to large galaxies and 
preferentially isolated dIrr (van den Bergh 1994) may suggest some kind of 
environmentally
driven evolution, either related to ram pressure sweeping (Lin \& Faber 
1983), or enhanced star formation that rapidly consumed the gas 
 in high density environments (Davies \& Phillips 1988; Lacey \& Silk 
1991).

Phoenix has the characteristics of a system in which a transition from
a dIrr galaxy to a dSph galaxy may be currently going on. It shows recent 
star formation in its central
part (500 to 100 Myr ago), but no HI has been observed centered in the
optical galaxy. Nevertheless, several clouds at different radial
velocities have been detected close to it. Carignan, Demers \& 
C\^ot\'e (1991) detected HI emission at $V_{\sun}=56$ \kms which was
clearly separated from a much larger scale component at $\sim$ 120 \kms 
associated with the Magellanic Stream. Oosterloo et
al. (1996) detected another smaller cloud at $V_{\sun}$=
--23 \kms, situated at 6$\arcmin$ SW from the center of the
galaxy. VLA observations by Young \& Lo (1997)
confirmed the existence of this last cloud close to the
optical galaxy, forming a curved structure that wraps around its SW
border. Subsequently, St-Germain et al (1999) reobserved a $2 \deg \times 
2\deg$ area around Phoenix with the Australia Telescope Compact Array, 
remapping all the previously observed components, plus one at 7 \kms
which is likely of Galactic origin. After a thorough discussion of all 
available evidence, these autors conclude that, due to its compact shape 
and internal velocity gradient, the amount of HI mass ($\simeq 10^5$ 
M$_\odot$) derived for the cloud if at the distance of the galaxy, and its 
location, almost overlapping the SW outskirts of Phoenix,  
the --23 \kms component is the one most likely associated with 
Phoenix.  It is 
interesting that the spatial variation of the young 
population with age in Phoenix is consistent with a possible 
self-propagation of the star formation over its disk from East to 
West (Mart\'\i nez-Delgado et al. 1999; see also Held, Savianne 
\& Momany 1999). Therefore, if the --23 \kms component
would prove to be associated with Phoenix, it would provide
tantalizing evidence that a burst of star formation may blow-out the 
gas from a dwarf galaxy. However, without a measurement of the radial 
velocity of the Phoenix stellar component, it is not possible to stablish 
a reliable connection between the stars and the HI gas seen near Phoenix 
in projection.

We present here the first measurement of the radial velocity of the stellar 
component in Phoenix, and discuss the possible association between the 
galaxy and the gas clouds in its neighbourhood in the sky. This paper is 
organized as follows: in 
Section~\ref{obs} we present the spectroscopic observations performed 
with FORS1 at VLT-ANTU. The derivation of the radial velocities of the 
Phoenix red giant branch (RGB) stars is described in~\ref{vel}. Finally,
the implications of the derived Phoenix radial velocity regarding the 
possible association to the dIrr of the different HI clouds detected toward 
it is discussed in Section~\ref{discussion}, based on the results of 
semi-analytical and numerical simulations of SNe explosions and ram-pressure
stripping scenarios. Our conclusions are summarized in 
Section~\ref{conclusions}.

\section{Observations} \label{obs}

Spectroscopy of a sample of candidate Phoenix stars and several 
velocity standard stars was obtained with 
FORS1 at VLT-ANTU on the night of September 14-15, 1999. The observing 
conditions were apparently photometric, but the seeing was mediocre,
about $\simeq$ 1-1.5\arcsec~ the whole night with some improvement only
at the end of the night, down to 0.7\arcsec.

We knew from past experience that, even for red giant stars, the best 
spectral range to
observe them for radial velocity measurements purposes is between about 
450-530 nm (Vogt et al. 1995). In this region, even for
relatively metal-poor stars, we obtain the highest density of spectral
lines, adequate continuum stellar flux, and low sky emission. We extended 
the spectral coverage some more to the red to include some sky lines, in
particular the [OI] 5577 line, to set the zero-point of the wavelenght 
callibration. In FORS1, the grism that covers that spectral region is  
GRIS600B. Its spectral resolution is R=815, which, with a resolution 
of 50\AA/mm and a CCD pixel size of 24$\mu$m, gave a 
dispersion of 1.2\AA/pixel.

We observed in multislit mode (MOS) to obtain spectroscopy of a sample of
stars in the field of Phoenix (Figure~\ref{mapa}). The usable field of view 
of $6\arcmin.8 \times 4\arcmin$ basically covered the main body of the 
galaxy. In two MOS setups 
(see Table~\ref{journal} and Figures~\ref{slitlet1} and~\ref{slitlet2}), 
we were able to employ 14 and 16 slitlets on stars that had a very high 
probability of membership as red giants in the galaxy, as selected from 
the Mart\'\i nez-Delgado et al (1999) photometry. A second star happened 
to be aligned with the selected target star
in two of the slitlets of the second set.  Figure~\ref{cmd} shows 
a Phoenix CMD with the observed stars marked.

We prepared approximate MOS slitlet 
configurations using the FIMS VLT software and the LCO image of Phoenix 
used in Mart\'\i nez-Delgado et al. (1999). This image did not allow
positioning the slitlets with enough precision to actually perform the 
observations, but it did allow to select optimal configurations of the 
19 slitlet set to secure maximum occupancy by the brightest red giants. The 
night before our observations, Dr. Iovino  (who was observing before us 
with FORS1) kindly took a 120 sec V FORS1 image of Phoenix for us, with 
which we were able to refine the slitlet positioning in such a way that 
no further adjustements were necessary during the target adquisition at 
the telescope.

Four  velocity standards of spectral type K0-K3 were 
observed a total of 7 times through the night. Because the exposure times 
necessary to obtain good S/N spectra of the standard stars were short to 
provide a good measurement of the position of the 5577 OI line, we took, 
immediately after each standard star measurement, a companion spectra of 
the sky, with exposure times of 300-600 sec. These companion spectra proved 
indeed very 
useful to obtain reliable corrections of the zero point of our wavelenght 
calibration.  Table~\ref{tabvstd} identifies the standard stars observed, and 
lists the velocity measurements. In addition, we took a 300 sec exposure 
of the twilight sky to be used as an extra, high S/N velocity template.

Arcs and screen flatfields were taken in the morning after our observations 
in each MOS setup. For practical considerations, the obtention of 
callibration arcs during 
the night is not supported with FORS1. This is because the whole operation 
of recording a wavelength calibration spectrum at night requires 
approximately 30 minutes. Because it involves
a reconfiguration of the telescope and of the instrument, it does 
not offer any significant advantage, compared to next morning daytime
calibrations, in terms of stability of the configuration with respect
to the one used for the observation of the scientific target. 
Therefore, the use of arc lamp spectra is restricted to 
defining the dispersion relation while zero point alignment is more 
accurately achieved using the night sky lines as references.

\section{Velocity measurements} \label{vel}

\subsection{Basic reduction procedures}

Because this is our first paper describing radial velocity measurements 
using FORS1 at the VLT, we will describe our reduction techniques in some 
detail. This section may be skipped  by any readers uninterested in the 
precise reduction procedure. We mainly used routines from the EXPORT 
version V2.11 of Sun/IRAF, with occasional use of MIDAS routines.

1) The data were taken in single amplifier mode and high gain 
(0.68 e$^-$/ADU and 5.03 e$^-$ read-out noise). We used IRAF 
to remove the bias level from the images, by using a set of 
bias frames taken in the same mode, plus the overscan region 
of each image. 

2) Screen flatfields were obtained in each slitlet configuration 
using the two available sets (blue and red filtered) of halogen 
flatfield lamps. Four flatfields were averaged for each configuration 
using IRAF, and subsequently normalized using NORM/MOS whithin the 
MIDAS package. This routine averages the rows  separately for each slitlet, 
smooths with a median filter (of 10 pixels here) and divides 
each row of the slitlet by the filtered average. The division of each 
object frame by the corresponding flatfield was performed using 
IRAF.

3) Three 1500 sec exposures were obtained in one of the slitlets 
configurations while two sets (1500 + 900 sec and 2$\times$1800 sec) 
were obtained in the second. No noticeable shifts were detected in 
either the spatial or the dispersion direction, and the spectra in each
set were averaged to increase S/N and eliminate cosmic rays.

4) Individual 2-D spectra were cut from the FORS1 image. From these, 
one-dimensional spectra were extracted  using 
the {\it noao.twodspec.apextract.apall} package in IRAF. At this point  
we also substracted the sky background by fitting it across the 
dispersion at either sides of the star spectrum. 
The substracted sky spectrum
was kept in one of the beams of the multispec format in order to be 
eventually used at a later phase to refine the wavelenght callibration 
using sky lines. 

5) He + HgCd lamp callibration spectra in each slitlet configuration 
were obtained the morning after the Phoenix and standard stars' 
observations, and were extracted, except for the background substraction, 
with identical parameters as the object spectra. 
We used {\it noao.onedspec.identify} to 
identify the lines in each callibration spectrum. Order 4 Chebyschev
polynomia were typically fitted to the position of 11-14 spectral lines 
and the derived transformations were applied to the program objects using the 
{\it noao.onedspec.dispcor} package.

6) The wavelenght callibration zero-point was checked and refined 
using the [OI] line at 5577.34 \AA. This line is present with high
S/N in all Phoenix long exposure spectra, but it is too faint in the 
radial velocity standard star spectra to provide a 
reliable measure of its position. For them, we used the longer exposure 
spectra obtained after shifting slightly the pointing of the telescope to
remove the standard star from the slitlet (see Table~\ref{journal}). The 
center of the 5577 line was obtained by fitting a gaussian profile using 
{\it noao.onedspec.splot}. The fact that there is some correlation between the 
position of the telescope and the amount and sign of the shift (see 
Table~\ref{shifts}) gives us 
confidence that we are indeed measuring an actual effect of telescope or 
instrument flexures on the spectra. The calculated $\delta \lambda$ shifts, 
typically within $\pm$ 0.4 \AA (Table~\ref{shifts}) were applied 
to both Phoenix and standard stars' images, by modifying the CRVAL1
header parameter, which provides the starting wavelength of each spectra.

7) The spectra were continuum-substracted using {\it noao.onedspec.continuum}. 
Finally, for the Phoenix spectra, we removed all pixels with values 
$\ge$ 40 or $\le$ --40 by 
replacing them by 0.0 using {\it images.imutil.imreplace} to get rid of any 
sky residuals and remaining cosmic rays.

\subsection{The Master Radial Velocity Template}\label{master}

We combined the standard star's observations plus the
twilight solar spectra into a single, high S/N, master template spectrum. 
To do this, 
we first transformed these spectra to a log $\lambda$ scale using 
{\it noao.onedspec.dispcor}. We then used the cross-correlation task 
{\it noao.rv.fxcor} to determine the relative velocities between each standard 
star spectrum and an arbitrarily chosen reference spectrum (HD 176047\#3).
The relative velocities ${\rm V_{rel}}$ were transformed to $\delta 
(\log \lambda)$ increments
as  $$\delta (\log \lambda)= {\rm -V_{rel}} \times \log e/c $$ 
and added to the 
CRVAL1 header parameter, which, in a logarithmic wavelenght scale, provides 
the logarithm of the starting spectrum wavelenght. Then, we combined 
the individual spectra into a single, high S/N template using 
{\it noao.onedspec.scombine}. By construction, the ``heliocentric velocity'' 
of this master template is that of HD 176047, i.e. --42.8 \kms. 


We then used the master template to obtain the {\it observed} radial 
velocity of the velocity standards. From the differences between
the observed and {\it true} heliocentric velocity, we can i) calculate 
an offset between the radial velocity template and the overall velocity 
system defined by the standards, which will be applied to the Phoenix 
velocity measurements to transform them to a {\it true} heliocentric system,
and ii) estimate the mean error with which we 
are determining the velocities of the standard stars. Because of the 
higher S/N of the standard star spectra, this will be
a lower limit of the error in the Phoenix velocity determinations. Only 
the standards for which we had a companion long exposure sky spectra to 
correct the zero-point of the wavelenght callibration have been used for this
purpose.  Table~\ref{tabvstd} lists the results of this comparison. The 
$\sigma_v$ values have been calculated as described in 
Section~\ref{secvelpho}. 
The $\sigma$-weighted mean difference between the true and observed 
heliocentric velocity is 
$V_{helio}-V_{obs}= 1.2$ with dispersion $\sigma_V$=1.7 \kms.

\subsection{Velocity of Phoenix giants} \label{secvelpho}

The radial velocities of the Phoenix giants were determined by 
cross-correlating each star with the master template described in 
section~\ref{master} using the IRAF task {\it noao.rv.fxcor}.  
Figure~\ref{figuraspec} shows the template spectra, an example of 
a typical Phoenix spectrum, and the corresponding cross-correlation 
function.

The $\sigma_v$ errors were determined, in the same way as described 
in Vogt et al. (1995) using six repeated measurements of two standard 
{\it plus} two repeated measurements of 6 Phoenix stars, listed in 
Table~\ref{repeat}. We assumed that the velocity errors can be written
as $\sigma_v=\alpha/(1+R)$, where $\alpha$ is a constant (Tonry \& 
Davis 1979). We then defined the $\chi^2$ statistic for the repeat 
measurements as $\sum (1+R_i)^2(v_i-<v>)^2/\alpha^2$ where $<v>$ 
refers to the mean velocity of the star corresponding to observation 
$i$. We then calculate the $\alpha$ value that gives a value of 
$\chi^2$ corresponding to a 50\% probability that we would exceed that
value of $\chi^2$ by chance (we will denote this value as $\chi^2_{50}$). 
With this definition, the adopted errors are equivalent to 1$\sigma$ Gaussian
errors. For this set of 18 observations of 8 stars with 10 degrees of freedom,
we have $\chi^2_{50} = 0.934$, from which $\alpha$=455.  

Table~\ref{tabcandidate} lists the measured individual heliocentric velocities 
for the observed Phoenix stars with the corresponding $R$ and $\sigma_v$ 
values. Only stars  with well defined cross-correlation peaks  have listed 
velocities. Some stars in Phoenix set\#2 have repeated measurements. Two 
of the observed stars turned out not to be RGB stars. Their spectra is briefly 
discussed in Appendix~A. 

Figure~\ref{dependencias} shows that there is no noticeable trend
in the measured velocities as a function of $V$ magnitude, $(V-I)$ color 
index, position angle PA, or galactocentric distance $r$ in Phoenix. The 
$\sigma_v$-weighted average of these velocities is --57 $\pm$ 6 \kms, 
the same as the median velocity.  The mean value of the Phoenix 
velocity remains stable within the errors if we eliminate from the 
average calculation stars with velocities $3\sigma$ above and below the 
mean. With this selection, we obtain  $V_{Phoenix-obs}=-53 \pm 6$ \kms. 
After correcting for the difference between the true and measured 
heliocentric velocity of the standard stars, we conclude that the {\it true} 
radial velocity of the stellar component in Phoenix is $V_{\odot}= -52 \pm 6$ 
\kms, which we will adopt in the remaining of this paper.

\section{Discussion.} \label{discussion}

\subsection{The stellar radial velocity of Phoenix and its association 
with the HI clouds.}

In their paper about the HI in the field of Phoenix, St-Germain at al (1999;
see their Figure~4)
discuss the identification of four components at heliocentric velocities 
--23, 7, 59 and 140 \kms. The components at 7 and 140 \kms are 
most likely of Galactic and Magellanic Stream origin respectively. The nature
of the other two components is unclear but St-Germain et al. (1999) conclude 
that the component at --23 \kms is most likely associated with Phoenix
because of i) its compact structure and partial overlap with the Phoenix 
stellar component, ii) the amount of HI mass inferred for the cloud at 
the Phoenix distance, which is similar to the mass of other dSph or dSph/dIrr
galaxies, iii) its velocity structure, unlike that of HVC,  which could 
indicate either rotation or ejection from Phoenix and iv) the location of 
the cloud west of Phoenix, consistent with the propagation from East to West 
of the most recent star formation event in the galaxy, as discussed
in Mart\'\i nez-Delgado et al. (1999). Our determination of the stellar radial
velocity of Phoenix $V_{\odot}=  -52 \pm 6$ \kms tends 
to support this particular association, if any, between HI in the field 
of Phoenix and the dwarf galaxy itself. The relatively large velocity 
difference of gas and stars, however, means that the association is not 
unquestionable and merits some discussion, as does the particular 
morphology and location of the HI cloud. 

In this section we will present the results of a series of semi-analytical 
and numerical simulations, under the assumption that the HI cloud 
at --23 \kms was once part of Phoenix and participated in the last star 
formation event observed in the galaxy about 100 Myr ago, which is the age 
of the younger HeB stars, found in the west region of Phoenix. We will 
consider two possible mechanisms that may have stripped Phoenix of its 
gas: gas ejection by SNe explosions and ram-pressure stripping. These models
allow us to set limits to the necessary conditions for this stripping to 
have taken place under a given mechanism, and we find out that both 
mechanisms may indeed have been responsible of the observed situation.  
Similar calculations under the assumption that the HI cloud at $+59$ \kms 
was once associated to Phoenix lead to unlikely boundary 
conditions (very large number of SNe required or large density of 
the intergalactic medium, IGM), 
which makes us conclude that this one is a chance superposition.

\subsubsection{Gas ejected by SNe explosions}

The possibility that an event of star formation in a small galaxy can 
actually blow out all or part of its gas, has been profusely discussed 
in the literature from the theoretical point of view (e.g. Mathews 
\& Baker 1971; Larson 1974; Dekel \& 
Silk 1986; Mac Low \& Ferrara 1999). This gas removal could eventually 
halt subsequent star formation and explain the low metallicity 
observed in dwarf galaxies and the relationship between their total mass 
and metallicity. Also, the HI mappings of some dwarf galaxies (Puche 
\& Westpfahl 1993) have shown that, in the smallest dwarf irregular 
galaxies, one large, slowly expanding shell usually dominates the 
interstellar medium.

In the case of Phoenix, we estimated the energy that should have been 
injected to the interstellar medium by SNe explosions if these were 
responsible for the ejection of the gas. We assumed that the HI cloud 
was ejected 100 Myr ago and projected in the opposite direction with 
respect to the observer (as indicated by the respective observed velocities 
--52 \kms and --23 \kms of Phoenix and HI cloud) and along the trajectory 
of Phoenix, in such a way that both observed velocities are affected by 
the same projection angle. We solved the equations of motion, considering 
the only effect of Phoenix's gravity force and imposing  
the projected distance between Phoenix and the HI cloud ($d \sim 0.65$ 
kpc) and their observed radial velocities (--52 \kms and --23 \kms, 
respectively) as boundary conditions. This calculation allows us to 
simultaneously estimate the energy necessary to eject the gas, 
$E_{SNe} \sim 2\times10^{51}$ erg and the most likely angle 
between the line of sight and the line connecting Phoenix and 
the HI cloud, $\alpha \sim 10$ degrees.  

To estimate the number of SNe necessary to inject this energy into the 
interstellar medium, we should estimate what fraction of the energy of 
a SNe is actually transferred as kinetic energy to the gas. Larson (1974) 
showed that the thermal energy available per SNe for driving a galactic 
wind in cases 
where SNe remnant cooling is important (as would be the case for dwarf 
galaxies) is of the order of $10^{50}$ erg, i.e. about 10\% the total 
energy output of the SNe. In the case of Phoenix, therefore, about 20 SNe 
would have been necessary to produce the inferred energy. 
By comparing the number of observed ~100 Myr old stars in the main sequence 
of Phoenix with the number predicted by synthetic CMDs, Mart\'\i nez-Delgado 
et al. (1999) calculate that the total mass involved in the burst of star 
formation that produced the younger stellar association in the western part 
of the galaxy is $1.8 \times 10^4 M_{\odot}$. Using the initial mass 
function of Kroupa, Tout \& Gilmore (1993), we derive that $\sim$ 50 stars 
more massive that $10 M_{\odot}$ should have been produced during that event,
providing a potential number of SNe large enough to produce the necessary
energy to have blown away the gas.

The most likely estimated angle $\alpha \sim 10$ degrees 
between the line of sight and the line connecting Phoenix and the HI cloud
means that both objects are in almost radial trajectories with respect 
to the line of sight. Therefore, their total velocities are very 
similar to the projected component, and the actual distance between 
Phoenix and the HI cloud is about 3.5 Kpc. 
Considering this distance and the relative velocities of the HI cloud 
and Phoenix, a Phoenix total mass $M_{Phoe-tot} \sim 4 
\times 10^8 M_{\odot}$ would be necessary if the HI cloud was to be 
gravitationally bound to Phoenix. This is about one order of magnitude 
larger than the total mass estimated for Phoenix, $\sim 3.3 \times 10^7 
M_{\odot}$ (Mateo 1998), which would imply that $(M/L)_{Phoe} \sim 400
M_{\odot}/L_{\odot}$, and indicates that indeed the amount of energy 
produced by the last burst of star formation would have been large 
enough to blow away the small amount of gas remaining in the galaxy after 
that particular episode of star formation. 

This model of SNe driven winds, even though very convincing from the 
energetic point if view, has the drawback of the regular appearance of 
the HI cloud, and the highly anisotropic geometry of the gas-stars 
configuration, with all the gas out in one direction in a single isolated
cloud. These are not a problem for the ram-pressure hypothesis 
discussed in the next section.

\subsubsection{Ram-pressure stripping} \label{ram}

The gas of the IGM produces a ram-pressure on 
the gas of an object (satellite) moving through it

\begin{equation}
P_{ram} \propto \rho_{IGM} v^2
\label{rampress}
\end{equation}

where $\rho_{IGM}$ is the gas density of the IGM and $v$ is the velocity
of the satellite (Gunn \& Gott 1972) relative to the 
IGM. When the pressure is strong enough, the 
gas of the satellite is slowed down and stripped from the satellite.
We have simulated a gas rich dwarf galaxy moving through a gaseous 
environment, using a semi-analytical approach to model the ram-pressure 
stripping (see G\'omez-Flechoso 2000 for details of the numerical 
models). We have analyzed the conditions for the HI cloud to be 
stripped by the IGM. We assumed an ambient medium of uniform density and
at rest with respect to the Galaxy, a peak of the gas density of the HI 
cloud of $4 \times 10^{19}$ atoms/cm$^2$ and the total luminosity of 
Phoenix of $L_B = 9.5 \times 10^5$ L$_{\odot}$
(St. Germain et al. 1999). Various $M/L$ ratios have been considered for 
the dwarf galaxy in our simulations. As in the
previous section, we reproduce the observational velocities 
and positions of Phoenix and the HI cloud, assuming
that (i) the gas has been stripped along the trajectory, and (ii)
the ram-pressure started 100 Myr ago. In this case, the forces involved in 
the trajectory of the cloud are the gravitational force 
of the dwarf galaxy and the ram-pressure stripping force
due to the IGM. The latter is produced by the ram-pressure given by eq.
(\ref{rampress}), taking into account that the velocity $v$ in the equation
is the velocity relative to the IGM, which is the Galactocentric 
velocity. The Galactocentric radial components of the velocity
of Phoenix and the HI cloud are $V_{R,Phoe}^{GC}=-144$ \kms, and
$V_{R,HI}^{GC}=-115$ \kms, respectively.
As in the calculation relative to the SNe in the previous section, in 
order to reproduce the projected distance between Phoenix and the HI 
cloud and their Galactocentric radial velocities, 
we estimate the angle, $\alpha$, between the 
line of sight and the trajectory of the satellite, and the gas density of 
the environment, $\rho_{IGM}$ (assuming it is nearly constant).
The best model corresponds to $\alpha \sim 21$ degrees, roughly 
independent of the $M/L$ ratio of Phoenix, and
$\rho_{IGM} \sim 2.0-3.0 \times 10^{-4}$ atoms/cm$^3$, for 
$(M/L)_{\rm Phoenix} = 2-20$ M$_{\odot}$/L$_{\odot}$.
This value of $\alpha$ implies that: (i) the orbit of Phoenix 
is quite eccentric, (ii)
the true distance between Phoenix and the HI cloud is about $1.8$ kpc,
and (iii) the total velocity of Phoenix and the HI cloud
are $V_{Phoe}^{GC} \simeq -154$ \kms and $V_{HI}^{GC} \simeq -123$ \kms.
The value of the gas density of the environment, $\rho_{IGM}= 2.0-3.0
\times 10^{-4}$ atoms/cm$^3$, obtained in this case is consistent 
with some estimates of the hot gas density of a Local Group halo (Suto et 
al. 1996; but see also Murali 2000; Mulchaey 2000 and references therein).

In Figure~\ref{proyec_vel_def} we show the results of the simulation: 
the gray levels refer to the predicted velocity of the gas (darker 
corresponds to more negative values of the velocity) and the white
contours correspond to heliocentric radial velocities --27, --24, --21 and 
--18 \kms. The black contours correspond to the predicted gas density 
(3.5, 3.0, 2.5, 2.0 and 1.5 $\times 10^{19} {\rm cm}^{-2}$). The 
structure of the gas in the simulations remains quite smooth as it is 
stripped from the satellite, keeping a distribution similar to the 
observed HI cloud. Note also that
a) a velocity gradient in the sense that the most extreme differences 
in velocity are seen farthest from the galaxy, and b) the velocity of 
the stripped gas is closer to the systemic velocity of the Milky Way 
than that of Phoenix are predicted by our model (see also Blitz \& 
Robishaw 2000), in agreement with the observed Phoenix parameters.

These arguments support the hypothesis of the ram-pressure as responsible of 
the gas stripping in Phoenix. However, it may demand Phoenix to be in its 
first approach to the Galactic Center, which may be possible if Phoenix is not 
bound to the MW or it is in a very long period orbit. Given its almost radial 
orbit, near the pericenter Phoenix must cross zones with a much larger gas 
density than in its current position, and at much larger velocity with respect 
to the IGM. This would imply a much larger ram-pressure stripping, to which the 
gaseous content of the galaxy could have only survived if it originally was
very massive and concentrated, unlike at the present time. 

There is at least a couple of theoretical works that consider tidal stripping 
of a gaseous satellite approaching to the Milky Way. They suggest that such 
a survival in the close approach may not be completely unrealistic. Mayer 
et al. (1999, 2001) show that low surface brightness 
satellites of a Milky Way-like galaxy, in eccentric orbits (as typical 
in hierarchical models of galaxy formation), may lose up to 90\% of their 
stellar mass due to tidal stripping. But, could any amount of gas remain 
still bound to the galaxy after one or more pericentric passages? Lin \& Murray
(1999) argue that ``around the dSph galaxies in the Galactic halo, their 
ejected unbound gas follows their orbits analogous to the Magellanic Stream. 
Gas density declines/enhances near peri/apogalacticon as the stream is 
stretched/compressed (by tidal forces). At apogalacticon, where the density 
is at a maximum and 
exposure to  the Galactic UV flux is at a minimum, the diffuse gas can recombine
to form atomic and molecular hydrogen and eventually stars''. To assess the 
plausibility of these different alternatives when ram-pressure is included in 
the case of Phoenix, it would 
be necessary to perform much more detailed simulations, which exceed the scope 
of the present paper. In any case, gas loss will be maximum at the pericenter, 
where both tidal stripping and ram-pressure forces are larger.  

\subsection{Comparison with observations of other dwarf galaxies}

In two cases, DDO 210 and LGS3
for which HI superimposed to the stellar component had been observed, Blitz 
\& Robishaw (2000) detect a stronger emission component offset from the stellar
component. They also find HI possibly associated with Leo I, whose position is 
also offset from the stellar component. Leo I and LGS3, for which an
optical velocity is also known, present interesting similarities with the
Phoenix case. In LGS3, HI cloud and galaxy differ in velocity by about
50 \kms, which implies that they are not gravitationally bound. In the 
case of Leo I, the velocity centroid of the emission that seems more closely 
associated with the galaxy (both spatially and kinematically) differs by 
about 30 \kms from the velocity of the stellar component. Apart from
the magnitude of the difference in velocity between gas and stars, Leo I, 
LGS3 and Phoenix are similar also in a) the morphology of the gas relative 
to the stars and in b) the time of the last events of star formation, 
which is of the order of a few hundred Myr in all of them (Phoenix: 
Mart\'\i nez-Delgado et al. 1999; Leo I: Gallart et al. 1999a,b; LGS3: 
Aparicio et al. 1997b). For Leo I and LGS3, Blitz \& Robishaw (2000) discuss
the possibility that the HI has been stripped by the ram-pressure of the hot
(Galactic or M31) halo gas. In spite of a distance to the Milky Way of almost twice 
that of Leo I, this seems a plausible explanation also for Phoenix, as we have 
discussed in Section~\ref{ram}. 

\subsection{Phoenix as a Milky Way satellite, and the mass of the Galaxy}

The minimum required mass of the Milky Way for Phoenix to be bound 
(based on the values of the Galactocentric velocity of
Phoenix and its distance to the Galactic center obtained above) 
is $M_{MW}(<450 {\rm kpc}) \ge 1.2 \times 10^{12}$ M$_{\odot}$. 
This value confortably fits within most current estimates of the Milky 
Way mass (see Zaritsky 1999 for a review). For example, Zaritsky et al. 
1989 conclude that  $M_{MW} = (1.3 \pm 0.2) \times 10^{12}$ from timing 
arguments involving Leo I and the Milky Way; on another hand, Anguita, 
Loyola \& Pedreros (2000) calculate, from a recent LMC proper motion 
measurement  that $M_{MW} = (3-4) \times 10^{12}$ out to 50 kpc, if the
LMC is to be bound to the Milky Way (see also  M\'endez et al. 1999).
We conclude, therefore, that Phoenix is very likely the outermost  
(known so far) Milky Way satellite.

\section{Summary and conclusions} \label{conclusions}

We present the first measurement of the radial velocity of the Phoenix 
stellar component, using FORS1 in MOS mode at the VLT UT1 (ANTU) 
telescope. We have obtained spectra of 31 RGB stars. Two additional 
stars for which we obtained spectra turned out to be a Carbon star and 
a bright supergiant star (see Appendix A). From the spectra 
of the RGB stars we derive an heliocentric optical radial velocity 
of Phoenix $V_{\odot}=-52\pm 6$ \kms. On the basis 
of this velocity, and taking into account the results of a series of
semi-analytical and numerical simulations, we discuss the possible 
association of two of the four HI clouds observed in the Phoenix vicinity, 
those at velocities --23 and $+59$ \kms which are not associated 
to either the Milky Way or the Magellanic Stream (St-Germain et al. 1999; 
Young \& Lo 1997; Oosterloo et al. 1996). 

We conclude that the characteristics (mass, distance, velocity relative to
Phoenix) of the HI cloud with heliocentric velocity --23 \kms 
are consistent with this gas having been associated with Phoenix in the 
past, until the last event of star formation in the galaxy, about 100 Myr 
ago. Two possible scenarios for the loss of this gas 
are discussed: its ejection by the energy released by 
the SNe produced in that last event of star formation, and a ram-pressure
stripping scenario.  The assumption that the HI cloud at $+59$ \kms 
was once associated to Phoenix lead to unlikely boundary 
conditions (very large number of SNe required or large density of the IGM), 
which makes us conclude that this one is a chance superposition.

We derive that the kinetic energy necessary to eject the gas is 
$E_{SNe} \sim 2 \times 10^{51}$ erg. The number of SNe necessary 
to transfer the inferred amount of kinetic energy to the gas cloud is 
consistent with the total mass involved in the last event of star formation
$\sim 100$ Myr ago by Mart\'\i nez-Delgado et al. (1999).
Considering the derived distance between Phoenix and the HI cloud and their
relative velocities, the Phoenix total mass necessary for the HI cloud to be 
gravitationally bound  is about an order of magnitude 
larger than the total mass estimated for Phoenix. This indicates that 
the amount of energy produced by the last burst of star formation would 
have been large enough to blow away the small amount of gas remaining 
in the galaxy after that particular episode of star formation. The main 
drawbacks of this SNe driven winds scenario is the regular appearance of 
the HI cloud and its anisotropic distribution with respect to the stellar 
component.  Another possibility is that the HI gas observed at --23 \kms
was stripped as a consequence of ram--pressure by the IGM. In our 
simulations, the structure of the gas remains quite smooth as it is 
stripped from Phoenix, keeping a distribution similar to that of the 
observed HI cloud.  

Since both the SNe and ram-pressure stripping scenarios for gas ejection 
in Phoenix are consistent with the observations, a combined scenario may be 
also plausible. One possibility is that supernovae resulting from a burst 
of star formation diluted the gas in Phoenix and increased its energy 
enough for it to be swept away by an already existing intergalactic 
wind.  Another possibility is that Phoenix may be crossing regions of
variable gas density, and that 500-100 Myr ago it ran into a relatively 
dense clump or stream of intergalactic gas that was responsible for 
the sweeping event that presumably terminated -or temporarily
halted- the star formation. 

Finally, we discuss the possibility that Phoenix may be a bound Milky Way 
satellite. The minimum required mass of the Milky Way for Phoenix to be bound  
is $M_{MW}(<450 {\rm kpc}) \ge 1.2 \times 10^{12}$ M$_{\odot}$. 
This value confortably fits within most current estimates of the Milky 
Way mass.

\acknowledgments

It is a pleasure to thank I. Perez, R. Guzman, E. Hardy and S. Zepf for very 
useful discussions on spectroscopic reductions, to R. Larson and L. Mayer 
for enjoyable conversations about possible scenarios for gas stripping, 
and for a critical reading of the manuscript. J. Mulchaey and S.G. Navarro 
pointed us to useful references on gas in galaxy groups and spectral 
classification, respectively. We also thank A. Iovino, 
for taking an image of Phoenix the night before our observations which 
allowed us to produce an accurate MOS setup. Support for this work was 
provided by Chilean CONICYT grant 1990638, NASA grant GO-5350-03-93A and by 
the IAC through grant PB3-94.

\appendix

\section{A young star and a carbon star spectroscopically
confirmed (serendipitously) in Phoenix}

One of the stars originally selected as a candidate RGB star, 
Phoenix set\#2-6 and Phoenix set\#2-10s3 (Table~\ref{tabcandidate}) 
turned out to be other types of stars.

Phoenix set\#2-6 (Figure~\ref{norgb}a) has the spectra typical of a Carbon 
star (see for example Cannon, Niss \& Norgaard-Nielsen 1980 for a
sample spectra and discussion of the features present in a C star spectrum). 
It is  one of the two C stars previously confirmed spectroscopically 
by Da Costa (1993). It is the brightest red star in our sample, located 
above the TRGB (identified in the CMD in Figure~\ref{cmd} with a solid square).

Phoenix set\#2-10s3 (Figure~\ref{norgb}b) was not in our original candidate 
list. It was included in the extreme of one of the slitlets, and turned out 
to be a young star. Its spectra clearly shows the series of Balmer lines in 
absorption, and at this low resolution, low S/N, we can classify it in the
range of spectral types B5--A5 (Jacoby, Hunter \& Christian 1984). Using
(m-M)$_0$=23.0 and $A_V$=0.06 for Phoenix (Mart\'\i nez-Delgado et al. 1999),
the absolute magnitude for this star, {\it if} it belongs to Phoenix, is 
$M_V$=-2.83 (Table~\ref{tabcandidate}), which would correspond to a bright 
giant star of luminosity class II, and probably early A spectral type 
(Landolt-B\"ornstein 1982). The position of this star in the CMD is 
consistent with a  blue-loop star with Z=0.001 and age 100 Myr, formed in 
the last event of star formation in Phoenix.

\newpage

\begin{figure}
\caption[]{Identification chart for the stars measured in Phoenix. The image is a 
120 sec. V exposure at the VLT. North is down and East it to the left.}
\label{mapa}
\end{figure}

\begin{figure}
\caption[]{MOS setup configuration for Phoenix set\#1 (Table~\ref{journal}) . FORS1 reference stars are circled. West is up and North is to the left.}
\label{slitlet1}
\end{figure}

\begin{figure}
\caption[]{MOS setup configuration for Phoenix set\#2 (Table~\ref{journal}). FORS1 reference stars are also marked. North is up and East is to the left}
\label{slitlet2}
\end{figure}

\begin{figure}
\begin{center}
\epsscale{0.8}
\plotone{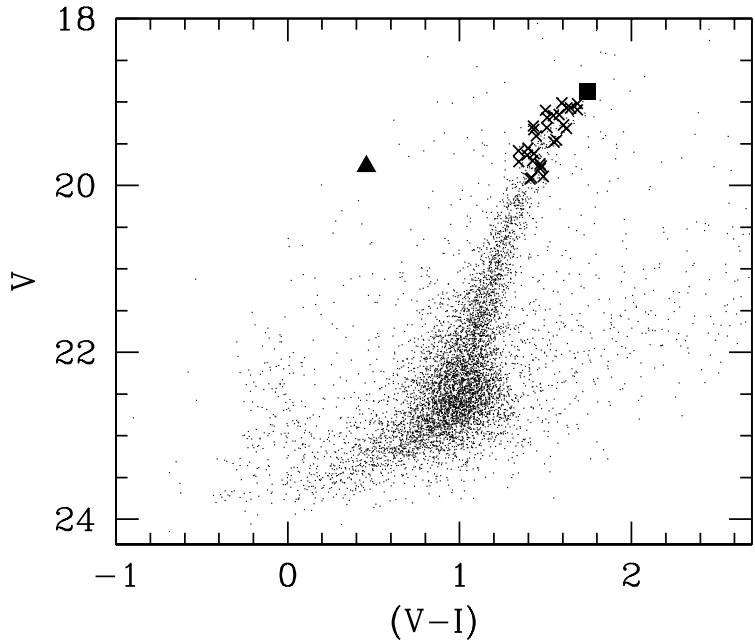}
\end{center} 
\caption[]{CMD of Phoenix with the measured stars near the tip of the 
red giant branch identified as crosses. The C star and the bright supergiant 
(late B- early A spectral type) star (see Appendix A) are identified with 
a filled square and a filled triangle respectively.}
\label{cmd}
\end{figure}

\begin{figure}
\caption[]{Top: Extracted spectrum of a typical Phoenix giant (Phoenix 
set\#1-14 in Table~\ref{tabcandidate}). Middle: Spectrum of the master 
radial velocity template. Bottom: The cross-correlation peak obtained from 
the top spectrum using the master template. The Tonry-Davis R value for this 
star is 11.6, which is the median value of all the stars considered in the 
Phoenix radial velocity derivation}
\label{figuraspec}
\end{figure}

\begin{figure}
\begin{center}
\epsscale{0.8}
\plotone{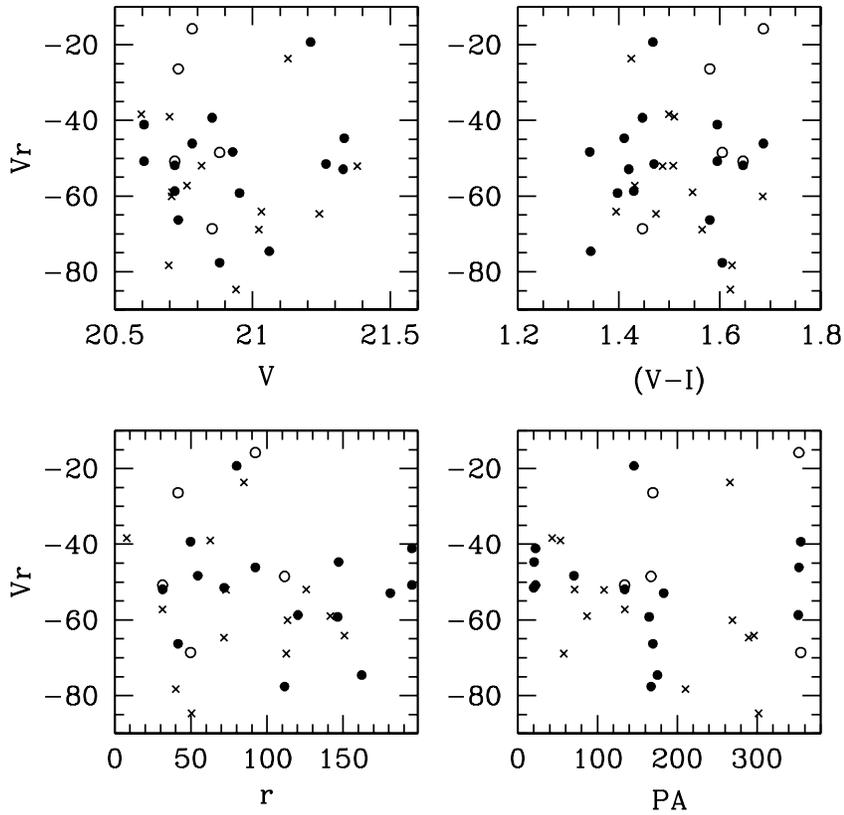}
\end{center} 
\caption[]{Radial velocities of Phoenix stars as a function of a number of 
parameters. Clockwise from top left: $V_r ~vs. V$ magnitude, $V_r ~vs. (V-I)$ 
color index, $V_r ~vs.$ position angle (in degrees), $V_r ~vs.$ galactocentric 
distance (in arcsec) within Phoenix. Crosses correspond to Phoenix set\#1, 
open circles to Phoenix set\#2.1 and solid circles to  Phoenix set\#2.2.  
Note that no trends in $V_r$ with respect any of these parameters, nor with 
the Phoenix set, exist.}
\label{dependencias}
\end{figure}

\begin{figure}
\begin{center}
\epsscale{0.8}
\plotone{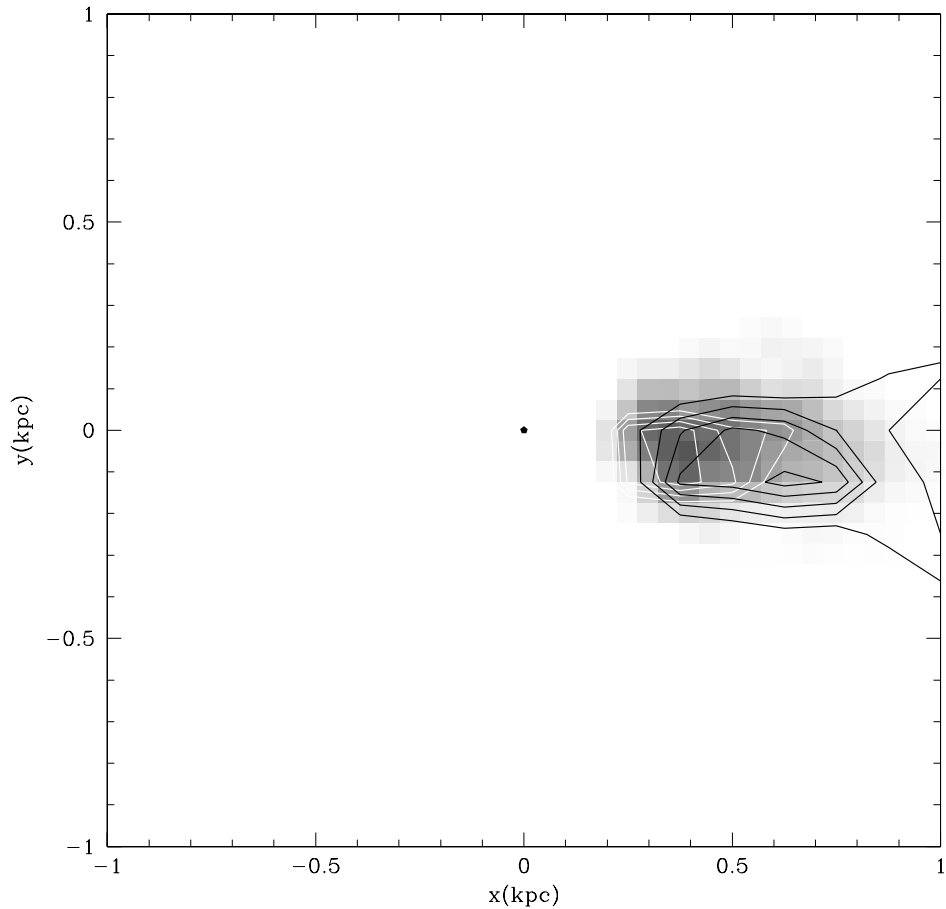}
\end{center} 
\caption[]{Results of the ram-pressure stripping simulation: gray levels 
refer to the predicted HI velocity (darker corresponds to more negative 
values of the velocity) and white contours correspond to heliocentric 
radial velocities --27, --24, --21 and --18 \kms. The black contours show 
the predicted gas density (3.5, 3.0, 2.5, 2.0 and 1.5 $\times 10^{19} 
{\rm cm}^{-2}$)}
\label{proyec_vel_def}
\end{figure}

\begin{figure}
\caption[]{Top: Spectrum of a Carbon star in Phoenix (Phoenix set\#2-6 in Table~\ref{tabcandidate}). Bottom: Spectrum of a young star in Phoenix 
(Phoenix set\#2-10s3), probably a bright giant (luminosity class II) 
of early A spectral type.}
\label{norgb}
\end{figure}

\newpage

\begin{table}
\caption {Observing log for the Phoenix and velocity standard stars}
\label{journal} 
\begin{center}
\begin{tabular}{llllr}
\tableline
\tableline
\noalign{\vspace{0.1 truecm}}
Object & $\alpha_{2000}$ & $\delta_{2000}$ & ARCHIVE FILE & Exp Time (s) \\
\noalign{\vspace{0.1 truecm}}
\tableline

HD176047\#1       & 18:59:44.12 & -34:28:14   & FORS.1999-09-14T23:44:30.357.fits &   10 \\
HD176047\#2       & 18:59:44.12 & -34:28:14   & FORS.1999-09-14T23:48:29.001.fits &   20 \\
HD176047$_{sky\#1}$& 18:59:45.73 & -34:28:14   & FORS.1999-09-14T23:51:50.492.fits &  300 \\
HD176047\#3       & 18:59:44.12 & -34:28:14   & FORS.1999-09-15T00:00:21.868.fits &  150  \\
HD176047$_{sky\#2}$& 18:59:45.73 & -34:28:14   & FORS.1999-09-15T00:06:18.654.fits &  600 \\
HD196983\#1       & 20:41:48.98 & -33:53:16   & FORS.1999-09-15T00:24:57.582.fits &  240 \\
HD196983$_{sky\#1}$& 20:41:47.44 & -33:53:16   & FORS.1999-09-15T00:34:39.793.fits &  300 \\
HD196983\#2       & 20:41:48.98 & -33:53:16   & FORS.1999-09-15T01:21:13.492.fits &  240 \\
HD196983\#3       & 20:41:48.98 & -33:53:16   & FORS.1999-09-15T01:28:43.540.fits &  120  \\
HD196983$_{sky\#2}$& 20:41:50.69 & -33:52:57   & FORS.1999-09-15T01:34:41.289.fits &  600\\
HD203638         & 21:24:07.99 & -20:50:59   & FORS.1999-09-15T03:39:29.409.fits &    5\\
HD203638$_{sky}$  & 21:24:09.27 & -20:50:59   & FORS.1999-09-15T03:42:16.512.fits &  600\\
Phoenix set\#1    & 01:51:04    & -44:26:05   & FORS.1999-09-15T06:03:47.856.fits & 1500\\
Phoenix set\#1    & 01:51:04    & -44:26:05   & FORS.1999-09-15T06:30:32.498.fits & 1500\\
Phoenix set\#1    & 01:51:04    & -44:26:05   & FORS.1999-09-15T06:57:17.324.fits & 1500\\
Phoenix set\#2.1    & 01:50:57    & -44:26:40   & FORS.1999-09-15T07:34:53.935.fits & 1500\\
Phoenix set\#2.1    & 01:50:57    & -44:26:40   & FORS.1999-09-15T08:01:38.649.fits &  900	\\	
Phoenix set\#2.2    & 01:50:57    & -44:26:40   & FORS.1999-09-15T08:18:01.062.fits & 1800\\
Phoenix set\#2.2    & 01:50:57    & -44:26:40   & FORS.1999-09-15T08:49:45.413.fits & 1800\\
cpd432597\#1      & 06:32:13.20 & -43:31:14   & FORS.1999-09-15T09:45:44.645.fits &   12\\
cpd432597\#2      & 06:32:13.20 & -43:31:14   & FORS.1999-09-15T09:48:03.264.fits &   48\\
\noalign{\vspace{0.1 truecm}}
\tableline
\tableline 
\end{tabular}
\end{center} 
\end{table}


\newpage

\begin{table}
\label{tabvstd} 
\caption{Radial velocity standards.}
\begin{center}
\begin{tabular}{lrrrrr}
\tableline
\tableline
\noalign{\vspace{0.1 truecm}}
Star & $V_{obs}$ & R & $\sigma_v$ & $V_{helio}$ &$V_{helio}-V_{obs}$  \\
\noalign{\vspace{0.1 truecm}}
\tableline
HD176047\#1	&       -40.8& 91.42&	4.9 & -42.8 &    -2.0 \\
HD176047\#2	&	-46.1&112.64& 4.0 & -42.8 &	  3.3 \\
HD176047\#3	&	-42.6&115.85&	3.0 & -42.8 &	  0.2 \\
HD196983\#1	&	  5.2& 62.45&	7.2 &  -9.3 &	-14.5 \\ 
HD196983\#2	&	-15.2& 98.44&	4.6 &  -9.3 &	  5.9 \\
HD196983\#3	&	-10.1& 79.65&	5.6 &  -9.3 &	  0.8 \\
HD203638	&	 10.6& 63.17&	7.1 &  21.9 &	 11.3 \\
\noalign{\vspace{0.1 truecm}}
\tableline
\tableline 
\end{tabular}
\tablenotetext{a}{The standard star radial velocities were taken from the following references: HD176047 and HD196983: Maurice et al. 1984; HD203638: Astronomical Almanac 1999}
\end{center} 
\end{table}

\newpage

\begin{table}
\caption{Dependence between wavelength zero-point shifts and telescope position. Values for the Phoenix sets are averages over all the slitlets}
\label{shifts} 
\begin{center}
\begin{tabular}{lrrrr}
\tableline
\tableline
\noalign{\vspace{0.1 truecm}}
Object & $\lambda_{obs} $ & 5577.34 - $\lambda_{obs}$ & Alt & Azim  \\
\noalign{\vspace{0.1 truecm}}
\tableline
HD176047$_{sky\#1}$ &	 5577.58 &	-0.24 & 79.6 & 342.0\\
HD176047$_{sky\#2}$ &    5577.45 &	-0.11 & 80.2 & 358.8\\
HD196983$_{sky\#1}$ &	 5577.70 &	-0.36 & 71.3 & 304.0\\
HD196983$_{sky\#2}$ &	 5577.45 &	-0.11 & 80.2 & 342.1\\
HD203638$_{sky}$    &    5577.15 &	+0.19 & 73.2 &  99.4\\
Phoenix set\#1     &    5577.66 &      -0.32 & 67.4 & 334.4\\
Phoenix set\#2.1   &    5577.05 &	+0.29 & 68.8 &  18.8\\
Phoenix set\#2.2   &    5577.08 &	+0.26 & 64.3 &  34.6\\
\noalign{\vspace{0.1 truecm}}
\tableline
\tableline 
\end{tabular}
\end{center} 
\end{table}

\newpage

\begin{table}
\caption{Summary of repeat velocity measurements}
\label{repeat} 
\begin{center}
\begin{tabular}{lrrrr}
\tableline
\tableline
\noalign{\vspace{0.1 truecm}}
Star & $V_{helio}$ & $R$ & $<V>$ & $V_{helio} - <v>$\\
\noalign{\vspace{0.1 truecm}}
\tableline
HD176047\#1 &	 -40.8 & 91.4& -43.2 &     2.4 \\
HD176047\#2 &	 -46.1 &112.6& -43.2 &	  -2.9 \\
HD176047\#3 &	 -42.6 &115.8& -43.2 &	   0.5 \\
HD196983\#1 &	   5.2 & 62.4&  -6.7 &    11.9 \\   
HD196983\#2 &	 -15.2 & 98.4&  -6.7 &	  -8.5 \\   
HD196983\#3 &	 -10.2 & 79.6&  -6.7 &	  -3.4 \\
Phoenix set\#2.1-5    &     -54.6 &  8.8& -66.3 &   11.61 \\
Phoenix set\#2.2-5    &     -77.9 & 13.4& -66.3 &  -11.61 \\	
Phoenix set\#2.1-8    &     -26.3 & 10.4& -47.3 &   21.01 \\	
Phoenix set\#2.2-8    &     -68.3 & 15.4& -47.3 &  -21.01 \\ 		 
Phoenix set\#2.1-9    &     -51.5 &  8.1& -52.3 &    0.80 \\		
Phoenix set\#2.2-9    &     -53.1 & 21.1& -52.3 &   -0.80 \\	 
Phoenix set\#2.1-12   &     -68.0 &  9.1& -53.9 &   -14.1 \\		
Phoenix set\#2.2-12   &     -39.9 & 13.9& -53.9 &	  14.1 \\	
Phoenix set\#2.1-14   &      -9.3 &  9.0& -28.3 &    19.0 \\	 	
Phoenix set\#2.2-14   &     -47.3 & 18.2& -28.3 &   -19.0 \\
Phoenix set\#2.1-15   &     -30.0 &  9.8& -44.8 &    14.8 \\ 
Phoenix set\#2.2-15   &	  -59.6 & 14.5& -44.8 &   -14.8 \\

\noalign{\vspace{0.1 truecm}}
\tableline
\tableline 
\end{tabular}
\end{center} 
\end{table}

\newpage

\begin{table}
\caption{Data of the spectroscopically observed stars in Phoenix, with radial velocities 
for the RGB stars with sufficient S/N to obtain a well defined cross-correlation peak.}
\label{tabcandidate} 
\tighten
\begin{center}
\begin{tabular}{lrrrrrrr}
\tableline
\tableline
\noalign{\vspace{0.1 truecm}}
Star & $\alpha_{2000}$& $\delta_{2000}$ & $V$ & $v_{helio_{obs}}$ & $R$ & $\sigma_v$ & Comment\\
\noalign{\vspace{0.1 truecm}}
\tableline
Phoenix set\#1-5 &  01:50:53.4 & -44:25:34.6& 21.03 & -64.1 &  10.1&  40.9 & \\
Phoenix set\#1-6 &  01:50:55.4 & -44:26:43.7& 20.71&  -60.1 &  16.9&  25.4 & \\
Phoenix set\#1-7 &  01:50:58.1 & -44:26:47.7& 21.13&  -23.7 &   9.8&  42.2 & \\
Phoenix set\#1-8 &  01:50:59.7 & -44:26:18.0& 21.24&  -64.7 &  17.4&  24.7 & \\
Phoenix set\#1-9 &  01:51:02.0 & -44:26:15.3& 20.94&  -84.7 &  18.9&  22.8 & \\
Phoenix set\#1-10 & 01:51:04.1 & -44:27:16.6& 20.70&  -78.3 &  16.3&  26.3 & \\
Phoenix set\#1-11 & 01:51:06.5 & -44:26:36.3& 20.60&  -38.4 &  27.1&  16.2 & \\
Phoenix set\#1-12 &   01:51:08.1 & -44:27:03.9& 20.76& -57.2 &  16.7&  25.6 & \\
Phoenix set\#1-13 &   01:51:10.7 & -44:26:04.8& 20.70& -39.0 &   9.7&  42.5 & \\
Phoenix set\#1-14 &   01:51:12.5 & -44:27:04.6& 21.38& -52.1 &  11.6&  36.1 & \\
Phoenix set\#1-15 &   01:51:14.9 & -44:25:41.8& 21.02& -68.9 &  21.3&  20.4 & \\
Phoenix set\#1-16 &   01:51:17.1 & -44:26:01.2& 20.82& -52.0 &   9.7&  42.4 & \\
Phoenix set\#1-17 &   01:51:19.2 & -44:26:34.3& 20.71&   --  &   -- &  --  & Low S/N\\
Phoenix set\#1-18 &   01:51:21.2 & -44:26:03.1& 21.05&   --  &   -- &  --  & Low S/N\\
Phoenix set\#2.2-2 &  01:51:05.1 & -44:29:42.6& 21.33& -52.9 &   8.8&  46.2 & \\
Phoenix set\#2.2-3 &  01:51:07.3 & -44:29:23.5& 21.06& -74.6 &   8.7&  47.0 & \\
Phoenix set\#2.2-4 &  01:51:09.6 & -44:29:03.1& 20.95& -59.2 &   7.4&  54.0 & \\
Phoenix set\#2.1-5 &  01:51:08.3 & -44:28:30.7& 20.88& -48.5 &   8.8&  46.2 & \\
Phoenix set\#2.2-5 &             &            &      & -77.6 &  13.4&  31.5 & \\
Phoenix set\#2.2-6 &  01:51:08.9 & -44:28:11.9& 20.62&  --   &   -- &  --   & C star\\
Phoenix set\#2.2-7 &  01:51:10.2 & -44:27:48.2& 21.21& -19.3 &  16.6&  25.9 & \\
Phoenix set\#2.1-8 &  01:51:06.7 & -44:27:22.8& 20.73& -26.4 &  10.5&  39.4 & \\
Phoenix set\#2.2-8 &             &            &      & -66.3 &  15.6&  27.3 & \\
Phoenix set\#2.1-9 &  01:51:08.1 & -44:27:03.9& 20.72& -50.8 &   8.2&  49.3 & \\
Phoenix set\#2.2-9 &             &            &      & -51.9 &  21.4&  20.3 & \\
Phoenix set\#2.2-10&  01:51:02.8 & -44:26:45.5& 21.03& -92.1 &  16.1&  26.7 & \\
Phoenix set\#2.2-10;s2&01:51:02.8& -44:26:41.9& 21.34& -67.0 &  14.9&  28.6 & \\
Phoenix set\#2.2-10;s3&01:51:02.7& -44:26:32.3& 20.23&   --  &   -- &   --  & bright supergiant \\
Phoenix set\#2.2-11 & 01:51:10.8 & -44:26:23.9& 20.93& -48.3 &   8.3&  48.9 & \\
Phoenix set\#2.2-11;s2&01:51:10.8& -44:26:20.5& 21.72& -88.6 &  13.8&  30.7 & \\
Phoenix set\#2.1-12 & 01:51:05.6 & -44:25:52.4& 20.85& -68.6 &   9.1&  45.1 & \\
Phoenix set\#2.2-12 &            &            &      & -39.3 &  13.8&  30.7 & \\
Phoenix set\#2.2-13 & 01:51:08.3 & -44:25:34.4& 21.27& -51.5 &   8.8&  46.3 & \\
Phoenix set\#2.1-14 & 01:51:04.9 & -44:25:10.5& 20.78& -15.8 &   9.2&  44.4 & \\
Phoenix set\#2.2-14 &            &            &      & -46.1 &  18.6&  23.3 & \\
Phoenix set\#2.2-15 & 01:51:04.4 & -44:24:42.9& 20.72& -58.7 &  14.4&  29.5 & \\
Phoenix set\#2.2-16 & 01:51:10.8 & -44:24:24.3& 21.33& -44.7 &   7.9&  51.3 & \\
Phoenix set\#2.1-18 & 01:51:12.9 & -44:23:41.4& 20.61& -41.1 &   7.7&  52.2 & \\
Phoenix set\#2.2-18 &            &            &      & -50.8 &   9.9&  41.7 & \\

\noalign{\vspace{0.1 truecm}}
\tableline
\tableline 
\end{tabular}
\end{center} 
\end{table}


\begin{references}

\reference{} Anguita, C., Loyola, P. \& Pedreros, M. 2000, \aj, 120, 845

\reference{} Aparicio, A., Gallart, C. \& Bertelli, G. 1997a, \aj, 114, 669

\reference{} Aparicio, A., Gallart, C. \& Bertelli, G. 1997b, \aj, 114, 680 

\reference{} Bingelli, B. 1993, in ``Panchromatic View of Galaxies'', eds.
G. Hensler, Ch. Theis \& J. Gallagher. \'Editions Fronti\`eres, Gif-sur-Yvette, 
p. 173 

\reference{} Blitz, L. \& Robishaw, T. 2000 \apj, submitted (astro-ph0001142)

\reference{} Carignan, C., Demers, S. \& C\^ot\'e, S. 1991, ApJ, 381, L13

\reference{} Carignan, C., Beaulieu, S., C\^ot\'e, S., Demers, S., Mateo. 
M. 1998, \aj, 116, 1690

\reference{} Cannon, R.D., Niss, B. \& Norgaard-Nielsen, H.U. 1980, MNRAS, 196,
Short Comm, 1.

\reference{} Da Costa, G.S. 1993, in `` Proceedings of the ESO/OHP 
Workshop on Dwarf Galaxies'', eds. G. Meylan \& P. Prugniel, ESO Conference and
Workshop Proceedings No. 49.

\reference{} Davies, J.I. \& Phillips, S. 1989, \apss, 157, 291

\reference{} Dekel, A. \& Silk, J. 1986, ApJ, 303, 39

\reference{}  Einasto, J., Saar, E., Kaasik, A. \& Chernin, A.D. 1974, 
Nature, 252, 111

\reference{} Ferrara, A. \& Tolstoy, E. 2000, \mnras, 313, 291

\reference{} Gallart, C., Freedman, W.L., Mateo, M., Chiosi, C., Thompson, 
I.B., Aparicio, A., Bertelli, G., Hodge, P., Lee, M.Y., Olsewszki, E.O., 
Saha, A., Stetson, P.B. \&  Suntzeff, N. 1999a, \apj, 514, 665 

\reference{} Gallart, C., Freedman, W.L., Aparicio, A., Bertelli, G. \& Chiosi,
C. 1999b, \aj, 118, 2245

\reference{} G\'omez-Flechoso, M.A. 2000, in preparation.

\reference{} Gunn, J.E. \& Gott, J.R. 1972, \apj, 176, 1

\reference{} Held, E.V., Saviane, I. \& Momany, Y. 1999, \aap, 345, 747

\reference{} Jacoby, G.H., Hunter, D.A. \& Christian, C.A. 1984, ApJS, 56, 257

\reference{} Kroupa, P., Tout, C.A. \& Gilmore, G. 1993, \mnras, 262, 545

\reference{} Lacey, C. \& Silk, J. 1991, \apj, 381, 14

\reference{} Landolt-B\"ornstein, Vol 2b. Eds. K. Schaifers \& H.H. Voigt, 
Springer-Verlag Berlin, Heidelberg, New York 1982

\reference{} Larson, R. 1974, \mnras, 169, 229

\reference{} Lin, D.N.C. \& Faber, S.M. 1983, ApJ, 266, L21

\reference{} Lin, D.N.C. \& Murray, S.D. 1998, in {\it Dwarf galaxies 
and Cosmology}, Eds. T.X. Thuan, C. Balkowski, V. Cayatte,
and J.T.T. Van, Eds Fronti\`eres.

\reference{} Lo, K.Y., Sargent, W.L.W. \& Young, K.,  1993, AJ, 106, 507

\reference{} Mac Low, M.-M. \& Ferrara, A. 19999, \apj, 513, 142

\reference{} Mart\'\i nez-Delgado, D., Gallart, C. \& Aparicio, A. 1999, 
\aj, 118, 862

\reference{} Mateo, M. 1998, \araa, 36, 435

\reference{} Mathews, W.G. \& Baker, J.C. 1971, \apj, 170, 241

\reference{} Maurice, E. et al. 1984, \aaps, 57, 275

\reference{} Mayer, L., Governato, F., Colpi, M., Moore, B., Quinn, T.R. 
\& Baugh, C.M. 1999, astro-ph/9903442 

\reference{} Mayer, L., Governato, F., Colpi, M., Moore, B., Quinn, T.R.,
Wadsley, J., Stadel, J. \& Lake, G.  2001, \apj, in press 

\reference{} M\'endez, R. A., Platais, I., Girard, T.M., Kozhurina-Platais, V. 
\& van Altena, W.F. 1999, \apj, 524, L39 

\reference{} Mulchaey, J.S. 2000, \araa, 38, 289

\reference{} Murali, C. 2000, \apj, 529, L81

\reference{} Oosterloo, T., Da Costa, G.S., Staveley-Smith, L. 1996, AJ, 
112, 1969
\reference{} Osterbrock, D. E., Fulbright, J.P., Martel, A.R., Keane, M.J.
\& Trager, S.C. 1996, \pasp, 108, 277

\reference{} Puche, D. \& Westpfahl, D. 1993, in {\it Proceedings of the 
ESO/OHP workshop on Dwarf Galaxies}, Eds. G. Meylan \& P. Prugniel.  ESO 
Conference and Workshop Proceedings No. 49, p. 273.

\reference{} Silk, J. Wyse, R.F.G. \& Shields, G.A. 1987, ApJ, 322, L59

\reference{} Smecker-Hane, T.A., Stetson, P.B., Hesser, J.E. \& Lehnert, 
M.D. 1994, AJ, 108, 507
\reference{} Stetson, P. B., Hesser, J.E. \& Smecker-Hane, T.A. 1998, \pasp, 110, 533

\reference{} St-Germain, J., Carignan, C., C\^ot\'e, S. \& Oosterloo, T. 1999, \aj, 1235, 1244

\reference{} Suto, Y., Makishima, K., Ishisaki, Y. \& Ogasaka, Y. 1996, \apj, 461, L33

\reference{} Tonry, J. \& Davis, M. 1979, \aj, 84, 1511

\reference{} van den Bergh, S. 1994, \apj, 428, 617

\reference{} Vogt, S.S., Mateo, M., Olszewski, E.W. \& Keane, M.K. 1995, \aj, 109, 151

\reference{} Young, L.M. \& Lo, K.Y. 1997, \apj, 490, 710

\reference{} Zaritsky, D. 1999, in ASP Conf. Ser. 165, The Third Stromlo Symp.: ``The Galactic Halo'', ed. B.K. Gibson, T.S. Axelrod \& M.E. Putnam (San Francisco: ASP), 34
 
\reference{} Zaritsky, D., Olszewski, E.W., Schommer, R.A., Peterson, R.C. \& Aaronson, M. 1989, \apj, 345, 759
 
\end{references}
\end{document}